\documentclass[article]{IEEEtran}

\usepackage[utf8]{inputenc}
\usepackage{subcaption}
\usepackage{graphicx, xcolor, url}

\title{Simplified State Storage Rent for EVM Blockchains}

\author{
\IEEEauthorblockN{Sergio Demian Lerner, Federico Jinich, Diego Masini, Shreemoy Mishra}\\
\{sergio, federico, dmasini, shreemoy\}@iovlabs.org\\
\IEEEauthorblockA{\today}
}
\begin{document}

\maketitle

\thispagestyle{plain}

\begin{abstract}
Uncontrolled growth of \textit{blockchain state} can adversely affect client performance, decentralization and security. Previous attempts to introduce \textit{duration-based state storage pricing} or \textit{storage rent} in Ethereum have stalled, partly because of complexity. We present a new approach with finer granularity to ``\textit{spread}'' rent payments across peers. Our proposal shifts the burden of state rent from `owners' or contracts  to transaction senders in a `quasi-random' manner. This proposal offers a simple path for initial adoption on Ethereum Virtual Machine (EVM) compatible chains, and serve as a foundation to address remaining challenges.
\end{abstract}


\section{Introduction}
In public blockchains, we distinguish between two broad sets of blockchain data. The first is the collection of cryptographically linked \textit{blocks},  which represents the core idea of blockchains as decentralized transaction timestamping servers. The second data set is the  \textit{current state} of the system - this is the information needed to actually execute a transaction.

\subsection{Blockchain State}
In Bitcoin, the state is the set of all unspent transaction outputs (UTXOs). Fully spent transaction outputs can be pruned to recover disk space. This is so simple that some users claim Bitcoin has no state. The state is richer in Turing complete smart contract platforms such as Ethereum. In \textit{Ethereum Virtual Machine} (EVM) compatible chains, state refers to data such as an account's (or smart contract's) \textit{balance}, \textit{nonce}, a contract's \textit{bytecode}, or the data contained in a contract's \textit{storage cells}. Transactions not only access state, they \textit{transform} it in the course of their execution. The \textit{stateroot} in an EVM chain's \textit{block header} is a cryptographic commitment of the cumulative, state-transforming, impact of all transactions in that block. 

Historical transaction data stored in blocks is not directly relevant for processing transactions. But this history can be used to recreate any past snapshot of the chain. Indeed, to re-execute an old transaction for validation  a client must first \textit{recreate} the state of the system -- exactly as it was -- when that transaction was initially processed and included in a block. This is the reason why new nodes takes a long time to synchronize with the network. For large chains such as Bitcoin and Ethereum, the initial block download and subsequent verification can take days on consumer grade hardware. This has lead to proposals to help jumpstart the process by offering new nodes a recent snapshot or \textit{checkpoint} of the state.

\subsection{The Problem with State Size Growth}
In a recent conference talk \cite{psz}, Péter Szilágyi, the lead developer of \textit{geth}\footnote{Go-ethereum or geth is the pervasive reference client implementation for Ethereum in the Go programming langauge.} remarked that Ethereum's state size growth places it on a ``\textit{potential death trajectory}''. While this may seem dramatic, such concerns are neither new nor exaggerated. Ethereum co-founder Vitalik Buterin \cite{vb0} and other core developers have been raising these concerns for a long time. Some consequences of continued expansion of the state include physical limits on how fast Ethereum (and EVM chains) can process transactions, and the inability of clients to sync with the blockchain.


In EVM compatible clients, state data are typically loaded to memory (RAM) using tree-like data structures called \textit{tries}. The concept of blockchain state trie was introduced and popularized in Ethereum's yellow paper \cite{gavin14}. The \textit{leaf nodes} in the trie contain actual state data for an account such as \textit{balance, nonce, storage root}, and \emph{code hash}.

On disk (or external storage drives), clients save the state trie in simple key-value datastores, most frequently using \textit{leveldb}. Keys represent the path or location of an account, or other state data, in the trie. Keys are constructed as Keccak hashes (as hash-prefixes in some cases) of data contained in a trie node, which helps avoid some types of DoS attacks. 

When executing transactions, we expect blockchain clients to spend most of their time manipulating relevant state data in memory, rather than reading from or writing to disk. In the early days of Ethereum, the protocol's architects were more concerned about the costs of computing hashes than disk input-output (IO)  operations. This tradeoff is part of the reason why they chose a \textit{hexary} trie over a binary one. All else equal, a binary trie is deeper, requiring more hash computations. Advances in computing power have reduced the cost of computing hashes. With rapid growth in state size, disk IO increasingly became the core bottleneck in executing transactions quickly.

If we want high transactions per second (TPS), then we must try and avoid going to disk too often. This is only possible if clients try to keep as much of the state as they can on RAM. Péter Szilágyi's talk \cite{psz} mentioned earlier has some striking observations. He estimates that the size of Ethereum's state is growing at the rate of about 20 GB per year. If we include the overhead of internal nodes and housekeeping, this blows up to an increase of around about 40 GB per year. 

It is not unusual for Ethereum core developers and professional node operators to run clients on machines with 64 GBs of RAM, which is far beyond typical user grade hardware. But even with such extraordinary amounts of RAM, there are no guarantees that we can avoid going to disk entirely. As the state grows faster relative to typical client memory, it becomes increasingly challenging for new clients to sync with the chain. Today, most Ethereum users and developers end up using API endpoints from hosted node service providers like Infura or Alchemy. This may become the fate of other EVM chains as well, which is not good for decentralization.

\section{Potential Solutions}

There are two main proposals to avert the worst consequences of massive growth in state size. The first is to introduce some form of state storage rent. This is the path we follow in this article. The second route is a set of ideas called \textit{state expiration}, where we give up on the notion of clients storing every piece of the current state, irrespective of how long it has remained unused. These proposals call for evicting ``stale'' information - which creates new problems. If some previously evicted or hibernated state is needed in the future, how can we load that information reliably? Each proposal for dealing with recovering `forgotten' data attempts to handle  the complex issues of doing so with some proof of validity. We do not delve deeper into state expiration any further. Our focus here is to present a proposal for duration based pricing as incentives to manage state size.

\subsection{Early Proposals}
The economic foundations for blockchain storage have never been very sound. Bitcoin introduced a very simple transaction pricing system of Satoshis per byte. This abstracted the true underlying cost of the P2P network such as computation, storage, bandwidth, or energy. In this model, the age of an UTXO had no impact on the fees while consuming it. This idea was carried over to newer blockchains including Ethereum and lead to user expectations of the form  ``pay once, store forever''. 

After a while, it became apparent that such poor economics would lead to unsustainable outcomes. Once the problem was apparent, proposals for time-based pricing (i.e. state rent) were inevitable. Ethereum core developers, including co-founder Vitalik Buterin, were early to recognize the value of storage rent. There is nothing novel about charging a fee for the duration of storing data. For instance, traditional cloud storage services such as AWS S3, Dropbox, or Google Drive have almost always had time-based (typically monthly) pricing from the start beyond \textit{free tiers}).

In traditional cloud storage, users "own" their data, and pay for storage based on space and time. Things are less clear in blockchain state. Who should pay rent for a popular library contract deployed by an anonymous user? Such fair accounting and attribution of rent costs has always been problematic. Many members of the community found the idea of rent incompatible with the `promise' of smart contracts, once deployed, could be called forever. There were several other complexities. For instance, if a preexisting contract was suddenly liable to pay for its storage, then its account balance would change in ways not anticipated at the time of deployment. 

The last strong push in Ethereum was lead by Alexey Akhunov \cite{alexey}. However, despite sincere effort and multiple iterations, the project did not gain broad community support. He directed attention to build a high performance ethereum client -- the \textit{Erigon} project (previously called \textit{turbo-geth}). Interestingly, Erigon is focused on reducing the number of IO operations per \texttt{SLOAD}, which is connected to this research on state access pricing.

Since the stalling of discussions related to storage rent, other alternatives have been proposed. These include the concepts of \textit{state expiration} or \textit{hibernation}, \textit{weak statelessness}, and \textit{regenesis}. Even if a blockchain takes some steps to reduce current state size, a new client syncing with the chain will still have to recreate all previous states for each transaction on the main chain. For Ethereum this is not even worth trying with ordinary consumer hardware and will get worse. Thus, it makes sense to move away from the current approach of having new clients go through a sync from genesis. 

However, if a new client starts syncing from a state of the system other than \textit{genesis}, they will need to get that state from somewhere, and this leads to trust issues. Therefore, some of the new proposals also require new proofs of validity i.e. \textit{witnesses}. At present, even these efforts appear to have stalled for over a year. This is partly because current attention is on ``\textit{the merge}``. The other reason is that Ethereum clients in operation are already high end machines, and those operating such nodes can deal with the additional RAM requirements in the near to medium term. However, for other EVM chains, the state may still be relatively small enough that introducing state rent can help maintain relative growth of state size and system RAM in ordinary user machines.

\subsection{Current Adoption Status}

Thus far, Solana protocol appears to be the only blockchain that has deployed a version of state rent on their main network \cite{solana}. Near protocol had implemented a version of storage rent and it was deployed in their testnet. However, they deprecated state rent before the launch of their mainnet. The concerns that lead to the deprecation largely mirror those faced by the Ethereum core developers. Near currently uses a model in which users stake tokens to gain storage rights \cite{near}.

Solana protocol expects storage costs to fall rapidly. This has two design consequences. The first is a two year horizon for their rent computations. This is driven largely by expectations about explicit (physical) storage costs being reduced by half each year. By contrast, our emphasis is on the \textit{implicit cost} of state size growth on the performance of nodes and decentralization. These are \emph{opportunity costs}, not physical costs. For instance, state size growth impacts ordinary users much more than block producers. Solana's emphasis is on block validators. Most end users are not really expected to run Solana nodes. By contrast, we are concerned about users who want to run their own EVM client nodes using consumer grade hardware on a home broadband connection.

\section{Our Proposal}
Rapid growth in state size is driven in large part by the lack of any pricing for storage duration. Every action by a user which increases state size imposes a negative externality on all other peers. Good engineering cannot make up entirely for poor economics. Such conflicts between private benefits and social costs are typically managed via taxes. Storage rent is just that - an externality tax. Unlike the costs of reading or writing to state (\texttt{SLOAD or SSTORE}), storage rent does not target the frequency of accessing state data. Instead it targets (the \textit{product} of) the \textit{size} and the \textit{duration} of storage.
 
Our main idea is very simple and generally applicable for all EVM chains. We note that this work is linked to an improvement proposal for the RSK blockchain \cite{rskip240} (also known as \textit{Rootstock}). For completeness, we present a brief note on RSK (following two paragraphs) which readers can skip without loss of continuity.

RSK is an EVM compatible Bitcoin sidechain \cite{lerner15}. It uses proof of work for consensus and is \textit{merged-mined} with Bitcoin by some of the largest mining pools (cumulatively representing over half of all hash power devoted to Bitcoin mining). RSK does not have an independent native currency, it uses Bitcoin. This is implemented via a \textit{federated 2 way peg} between Bitcoin and the native coin RBTC. The monetary supply of RBTC is determined by the amount of Bitcoin locked to the (multisig) federation address via \textit{peg-in} transactions on Bitcoin.  Like most federated pegs, the interesting bit is the \textit{peg-out} from the sidechain back to Bitcoin.  The Federation members do not control the signing keys, which are within hardware security modules (HSMs). These modules sign peg-out transactions only when instructed by a bridge smart contract.

Bitcoin miners who merged-mine RSK only earn transaction fees. There cannot be a block subsidy, since the network cannot create new RBTC without corresponding deposits of Bitcoin. Implementing storage rent can increase these fees, contributing to merged-mining rewards. This increase in block producer fees from storage rent holds for proof of stake chains as well \textit{e.g}, Solana.  

\subsection{Granularity of Rent}
In this proposal, rent tracking and computation is at the level of the \textit{individual pieces} of state data. For most EVM chains this an individual, value-containing (leaf) node, in the state trie. Ethereum has separate tries for accounts and storage. However, for simplicity we refer to a single state trie\footnote{The RSK blockchain actually has a single state trie that contains account and contract data as well as storage cells. Moreover, unlike Ethereum's hexary trie, RSK uses a binary tree structure.}.  

This granularity is different from earlier state rent proposals for Ethereum where rent computation was at the level of an account, e.g. all of the storage consumed by a contract. We adopt a much finer granularity. We track \textit{rent payment status} for each node in the trie by adding a timestamp for the last time rent was collected for that node. This can be the timestamp of the block in which the last rent update was made. The interpretation is that the node \textit{owes rent} since the timestamp.

\subsubsection*{Size}
The obvious unit for size for rent computations is \textit{bytes used}. However, when computing rent for a trie node, we should account for the overhead of storing information, including the key. For instance, even though a storage cell may contain 32 bytes of data, the effective size is closer to 100 bytes as there is an overhead of about 64 bytes. The same holds for a node containing account information. Contract bytecode is obviously much larger and the overhead has small impact.

\subsubsection*{Duration}
The duration of state storage can be measured either in units of time (\textit{e.g.} unix seconds since last rent update) or a chain may use difference in block height (since last update). We think seconds is more natural and is similar to that in previous proposals. 

\subsubsection*{Rental Rate}
Rent is collected in gas, in addition to regular transaction fees. Given the granularity, a \textit{rental rate} $R$ will have units of \textit{gas per byte per second}. In our proposal for the RSK network \cite{rskip240} we use a value $R = 1/2^{21} gas/byte/sec$. This value corresponds to an annual rent of about 2500 gas for a storage cell or an account. The rental rate is \textit{ad hoc} and cannot be determined accurately without an economic model of how a public EVM chain values state storage and the costs of disk IO. This is unlike setting the gas cost schedule for various EVM \texttt{opcode}s which can be linked to client performance benchmarks \textit{e.g.}, nanoseconds of CPU time under standardized test conditions.

\textbf{Illustrative Example:} Consider a transfer of some DAI (or any ERC20) tokens from Alice to Bob. Some of the state trie nodes touched by this transaction are: Alice's  account node (transaction sender), the trie nodes containing the DAI smart contract bytecode, the storage cells containing application parameters, the cells containing the mapping of addresses to token balances, and finally the storage cells that contain Alice's (token sender) and Bob's (token recipient) DAI balances. Most of these are trie reads with a few trie writes.

\subsection{Attribution of Rent}
Attribution asks ``\textit{Who pays state storage rent and for what}?'' In our proposal, we require the sender of a transaction to be \textit{partly responsible} for paying a portion of outstanding rent for each part of the state that their transaction reads from or writes to. As will become clear, we expect a very small fraction of transactions to trigger any rent payments. This is due to parameters for rent caps and collection thresholds (explained below).

The account-based tracking and collection of rent of earlier rent proposals has an implicit notion of \textit{ownership}. We think that this notion of data ownership is at odds with the idea of a public blockchain. We treat all state data as collectively owned, and therefore the cost should be distributed across users in a collective sense. However, collective ownership does not mean equal allocation of dues. Instead, we distribute the burden of rent based on who is consuming the data - i.e. a transaction sender.

For our illustrative example, in addition to the usual gas amount for transferring some DAI to Bob, Alice (as transaction sender) \textit{may have to pay} some storage rent for the trie nodes touched by the transaction. If the amount of rent due for any nodes is above the collection triggers, then all such rent (subject to caps) will be deducted from the transaction's gas limit.

\subsection{Rent Tracking and Computation}
As a transaction is executed, we keep track of all trie reads and writes. At the end of the execution, we check the rent timestamps for all the trie keys accessed (reads and writes), and compute the amount of rent due. The total amount of rent to be collected is consumed from the transaction's \textit{gaslimit}. Unlike execution gas, which is consumed on the go, we think it is better to collect rent at the end of transaction execution. Attempting to consume rent on the go can lead to breaking changes. For example, any contract calls that limit the amount of gas passed to the call can potentially break if we attempt to consume rent (from the restricted budget) for the trie nodes accessed by the internal transaction.

\subsection{Caps and Thresholds}
Updating a rent timestamp involves a trie put operation, which is costly. So we avoid collecting insignificant amounts in rent. Thus, for any trie node touched by a transaction, we should collect rent only if it exceeds the cost of updating rent timestamps. For example, we can set this minimum amount to trigger a collection at 5000 gas, which is the cost of a write operation (\textit{e.g.} cost of a storage cell update). Of course, if a new value has to be written anyway, \textit{e.g.} updating  an account or token balance, then the threshold for collecting rent can be a smaller (e.g. 1000 gas). Thus, we have lower thresholds for trie writes (e.g. \texttt{SSTORE}) and larger collection triggers for pure trie reads (e.g. \texttt{SLOAD}). 

If a part of the state is unused for a long time  (\textit{e.g.} an old library), it may have a large amount of rent due. We should not place the burden of large rent payments on the first user to access this unit of the state. Therefore, we place limits on how much rent can be collected for a trie node in a single transaction. For instance, we can set the cap at 10,000 gas per trie node (per transaction).

If the amount of rent due for a trie node exceeds the cap, then the timestamp cannot be advanced all the way as the current block timestamp. In such cases, the timestamp is advanced in a manner consistent with the node's size and the amount of rent collected (\textit{i.e} the cap).

\textbf{Quasi-random rent collection:} The existence of caps and thresholds implies rent collection is state dependent. The same transaction sent at a different time can lead to different rent computations. This is how rent payments are `randomly' distributed across users. The probability distribution (size and likelihood) of any user paying storage rent will depend on the type of user. Those who engage more frequently with decentralized applications, and with a wider variety of applications, are more likely to pay state rent than other users with limited participation. The quasi-randomness in rent payments can make gas estimation more challenging. However, we believe this is something wallet developers can adapt to and optimize.

\subsection{Internal Nodes, State Expiration and Caching}

Rent is only collected for value containing (typically, leaf) nodes in the state trie. However, we propose that rent timestamps should also be included in internal nodes as well. The timestamp for any internal node should be updated to match that of the most recently updated child node. Doing so can have several benefits. For instance, if an EVM chain decides to implement \textit{state expiration}, or \textit{hibernation}, then they can use rent timestamps to determine which parts of the trie to evict from state.

At present, we are not convinced about the soundness of proposals for state expiration and recovery for EVM chains. Our view is that storing data on drives is not that expensive (and is getting cheaper). Our attention should be on the limitations of RAM and disk IO. In this regard, we think EVM chains can use rent timestamps to create different levels of \textit{cache hierarchies} without ever deleting or evicting parts of the state. Instead, individual client operators can decide how to organize their disk storage and use rent timestamps to decide how far (from RAM) some parts of the state should be. For instance, the parts of state with the oldest timestamps can be stored in the slowest storage location. Portions of the state with recent timestamps can be on RAM. Each client can set a local policy. This will incentivize users and app developers to keep rent status fresh, so their transactions execute faster.

We should point out that our suggestion to use timestamps for caching is in contrast to a position adopted by Solana protocol. In that system, every part of the state is checked for rent status at the end of each epoch. There is no concept of dormancy or not paying rent during an epoch. This is starkly different from our proposal where a part of the state may not be touched for years. In our proposal, it is okay not to collect rent on a regular basis for every part of the state. We think it is fine for such data to just stay on disk without any consequence on RAM capacities or disk IO. This is also why we think that it safe to stop accumulating rent for a trie node once it has not been touched for some period of time, e.g. three years. There is no strong economic reason to keep accumulating rent forever.

\subsection{Interaction with Refunds}


At present, the only economic incentive for reducing state storage comes from \textit{refunds} for clearing storage cells and deleting (self-destructing) deployed contracts. However, refunds have proved problematic in the past. There is not much evidence that they encourage good state hygiene. On the contrary, refunds have been blamed for speculative hoarding of storage e.g. \textit{gas tokens}. This is why Ethereum reduced the refunds for storage cells. Storage Rent does not reduce state size by directly. That would be the case if rent is combined with some kind of state expiration policy. However, we believe rent can work \textit{in concert} with gas refunds and provide additional incentive to reduce storage.

\subsection{Reverts, Deletions and DoS Attacks}
Consider a transaction in which a contract loads hundreds of storage cells and then self destructs. In this case, the contract and all its storage are removed from state. Should we compute and collect storage rent in such cases? While it may seem like collecting rent is not worth the effort, we have to collect the rent due, up to the individual caps. Otherwise such patterns can be used for DoS attacks.

Another pattern that can be used for DoS is for an attacker to attempt to load parts of the state which they know not to exist. Even if these keys are not part of the real state, such lookups force clients to search on disk. One way to discourage such attacks is to raise the cost of trie reads. This is indeed the path followed by several improvement proposals in Ethereum including EIP-150,EIP-1884, and EIP-2929. Storage rent provides an additional way to deal with such attacks without altering the pricing of EVM \texttt{opcode}s. If a key that a user searched for does not exist, than we can collect a penalty as part of the rent mechanism. This can be made something like the equivalent of one year's rent for a storage cell. Arguably, such penalties will impose costs on legitimate use cases, and this can influence the choice of the penalty.

A proposal for storage rent must also consider what happens to the rent status of parts of the trie that are accessed by a transaction which gets reverted (including internal transactions). Since the transaction is reverted, no state changes will be recorded. However, in such situations we must still collect a fraction of the computed rent dues (\textit{e.g.} 25 percent)] for these state changes that are \textit{rolled back}. The amount collected is in part compensation for rent related computation, and also as deterrence against some DoS attacks. 

\section{Pricing of Rent}
As noted in Buterin \cite{vb0}, one of the primary challenges in deploying storage rent is how to determine the pricing. The gas cost schedule for various EVM opcodes is (loosely) proportional to the resources consumed. These costs get scaled by the \textit{gasPrice}, which ultimately is determined by gas fee market. Setting an appropriate level for storage rent will require experimentation and tinkering.

For storage rent to have any impact on state size growth or IO operations, the price must be economically significant. Measuring the impact requires price experimentation. Estimating users' and developers' response to rent will be hard and time consuming. Since rent is about economic incentives, deploying storage rent on some testnets will not be very informative. Testnet is fine for testing the engineering side of rent tracking, client performance, wallet integrations, and user experience. But the economic impact will require long term testing in a mainnet setting.

Suppose there is a proliferation of L2 chains that are essentially forks of geth. If some of them implement storage rent while others do not, then that could serve as a testing ground. However, even then, it would be hard to control for the multiplicity of factors that affect user growth and adoption across seemingly similar networks. Therefore, while plausible, we do not expect such experiments as a promising path. 

The only alternative is to have price experimentation within a single chain.  Introducing storage rent will require a network upgrade (a \textit{hard fork}). It may seem that performing any price experimentation may also hard forks to alter the parameters for rent. However, it is impractical to think about deliberately introducing hard forks in stable networks for price experimentation. There are ways to get around this restriction. The first, is to rely on natural variations in gasPrice and also coin prices. A more intricate manner of performing price experiments (closer to randomized trials) without hard forks is to include the experimental structures directly inside the client. For example, we can include deterministic or randomly scheduled \textit{rent-holidays}. These could be periods during which rent collections can be paused, or discounted in some manner. Such experiments can help recalibrate the rental rate and measure user responses, which in turn can lead to more effective pricing of state storage duration.    

\section{Conclusion}

Storage rent cannot solve all problems associated with state size growth. However, it is one of very few tools that utilize a microeconomics approach (of using externality pricing) to strike a balance between private  benefits and social costs. Our proposal is simple, yet highly stylized in some ways. For example, rent payments cannot always be estimated or predicted with precision. This will create some new headaches for wallet developers. However, these problems are largely manageable and unlike previous proposals, there are no breaking changes for preexisting contracts. It avoids the complications of state expiration or regenesis, but can work alongside them.

At present, only Solana has implemented storage rent. Introducing rent in EVM chains will require a hard fork. However, this is not a reason to avoid it. Ethereum core developers have repeatedly demonstrated their ability of introducing complex changes through open debate, community discussions, and thorough testing. The complexity of implementing storage rent is relatively mild in comparison to something like the recently activated "merge" that switched Ethereum's consensus mechanism from \textit{proof of work} to \textit{proof of stake}.

Part of the challenge in introducing rent is cultural, not technical. If Ethereum were to adopt rent, most other EVM chains will do so as well. If Ethereum does not adopt it, then most EVM chains will remain hesitant. However, it is entirely plausible that some chains experiencing rapid growth may be compelled to act sooner than Ethereum. As noted in \cite{psz}, some EVM chains e.g. BSC, dialed back from operating at very high scale with block sizes several times larger than Ethereum's. Perhaps they were able to foresee the problem of unmanaged state size growth. Perhaps some of these chains will join RSK in exploring state rent as one part of a sustainable approach to decentralized computation.

\bibliographystyle{IEEEtran}
\bibliography{biblio}

\end{document}